\documentclass[manuscript,screen]{acmart}
\usepackage{multirow}
\usepackage{arydshln}
\AtBeginDocument{%
  }

\acmSubmissionID{5739}

\begin{document}

% \title[Privacy-Preserving Student Stress Recognition through GPS Encoding and Activity-Based Features ]{Privacy-Preserving Student Stress Recognition through GPS Encoding and Academic-Based Features}

\title[From Coordinates to Context: An LLM-Bootstrapped Semantic Encoding Framework for Privacy-Preserving Mobile Sensing Stress Recognition]{From Coordinates to Context: An LLM-Bootstrapped Semantic Encoding Framework for Privacy-Preserving Mobile Sensing Stress Recognition}
%  in Inertial Sensing Data


\author{Hoang Khang Phan}
\email{khang.phan2411@hcmut.edu.vn}
\orcid{0009-0007-1578-0977}
\affiliation{%
\department{Department of Biomedical Engineering}
  \institution{Ho Chi Minh city University of Technology, Vietnam National University-Ho Chi Minh city}
  \city{Ho Chi Minh city}
  \country{Vietnam}
}

\author{Nhat Tan Le}

\email{lenhattan@hcmut.edu.vn}
\orcid{0000-0002-2738-6607}
% \authornote{Corresponding author.}
\affiliation{%
\department{Department of Biomedical Engineering}
  \institution{Ho Chi Minh-city University of Technology, Vietnam National University-Ho Chi Minh city}
  \city{Ho Chi Minh city}
  \country{Vietnam}
}

% \renewcommand{\shortauthors}{Hoang Khang Phan et al.}

%%%% The abstract is a short summary of the work to be presented in the
\begin{abstract}

Psychological stress is a widespread issue that significantly impacts student well-being and academic performance. Effective remote stress recognition is crucial, yet existing methods often rely on wearable devices or GPS-based clustering techniques that pose privacy risks and lack of human understandable explanations. 
In this study, we introduce a novel, end-to-end privacy-enhanced framework for semantic location encoding using a self-hosted OSM engine and an LLM-bootstrapped static map for human-friendly feature extraction, and pave a pathway for privacy-aware location data transformation for dataset sharing. We rigorously quantify the privacy-utility-explainability trilemma and demonstrate (via LOSO validation) that our Privacy-Aware (PA) model achieves robust privacy protection without being statistically distinguishable in stress recognition performance from a non-private model. Model explanation analysis highlights that our extracted features, which are user-friendly features, match with psychological literature about stress. In addition, an ablation study on the GeoLife dataset also demonstrates that our privacy framework improves privacy by 2-3 times compared to a non-privacy-aware approach. This suggests that our system can be utilized for the next generation of GPS transformations in open-source datasets for future researchers.
\end{abstract}


\begin{CCSXML}
<ccs2012>
   <concept>
       <concept_id>10002978.10003029.10003032</concept_id>
       <concept_desc>Security and privacy~Social aspects of security and privacy</concept_desc>
       <concept_significance>300</concept_significance>
       </concept>
   <concept>
       <concept_id>10003120.10003138.10003140</concept_id>
       <concept_desc>Human-centered computing~Ubiquitous and mobile computing systems and tools</concept_desc>
       <concept_significance>500</concept_significance>
       </concept>
   <concept>
       <concept_id>10010147.10010257.10010258.10010259</concept_id>
       <concept_desc>Computing methodologies~Supervised learning</concept_desc>
       <concept_significance>500</concept_significance>
       </concept>
 </ccs2012>
\end{CCSXML}

\ccsdesc[300]{Security and privacy~Social aspects of security and privacy}
\ccsdesc[500]{Human-centered computing~Ubiquitous and mobile computing systems and tools}
\ccsdesc[500]{Computing methodologies~Supervised learning}

\keywords{Stress Recognition; Student Life; GPS; Machine Learning; Feature Extraction}

% \received{}
% \received[revised]{}
% \received[accepted]{}

\maketitle

\section{Introduction}
In a fast-paced modern society, psychological stress stemming from heightened professional, academic, and social demands has become a pervasive epidemic. Left unmanaged, chronic stress diminishes productivity and poses severe risks to long-term well-being and longevity. In 2018, a YouGov survey found that over 70\% of people reported feeling so stressed that they were overwhelmed or unable to cope \cite{mentalhealthfoundation_stress_statistics}, and recent data indicates that 75\% of high school students experience constant stress \cite{transformingeducation_student_stress_statistics}. Consequently, there is an urgent need for accessible, scalable, and responsible AI-driven tools capable of identifying early signs of distress without compromising user privacy.

Smartphones, ubiquitous in modern life, offer unprecedented potential for continuous, passive mental health monitoring \cite{acikmese2019prediction,student_life4,student_life2,dasilva2019correlates}. Among the various modalities of smartphone sensor data, location tracking provides arguably the most valuable insights into human behavioral patterns \cite{hasan2013understanding,sila2016analysis}. Mobility features, such as time spent at work, travel duration, or engagement in recreational activities, serve as profound indicators of an individual's emotional state \cite{student_life2}. For instance, an active effort to mitigate stress may manifest as reduced time at the workplace and increased duration in leisure or shopping environments. 

\textbf{However, the continuous collection and secondary sharing of high-resolution Global Positioning System (GPS) trajectories present profound privacy vulnerabilities.} To advance the field of ubiquitous computing, researchers must be able to collect and share longitudinal datasets; yet, raw location data is highly susceptible to re-identification attacks (see Section \ref{ablation_study}). To circumvent this, existing public datasets and remote stress recognition systems heavily depend on GPS-based spatial clustering (e.g., extracting "Time at Top 1 Location") or noise-injection techniques \cite{student_life4,student_life3}. 

Crucially, while mathematical privacy models like Geo-Indistinguishability exist, \textbf{they fundamentally conflict with the Human-Computer Interaction (HCI) requirement for explainability}. When data stewards obfuscate coordinates into abstract spatial clusters prior to dataset release, they destroy the inherent semantic meaning of the locations. Consequently, downstream researchers are forced to utilize opaque, black-box deep learning models that cannot return human-sensible explanations for their stress assessments. To provide clinically actionable mental health insights, a system must be able to report \textbf{why} stress is occurring (e.g., 'A 40\% drop in recreational time'), a semantic clarity that both clustering and noise-injection inherently destroy.

To address this critical gap, this study introduces a novel GPS encoding framework designed as a standardized transformation protocol for privacy-aware location data collection and sharing. Rather than relying on abstract geometric clusters or invasive raw coordinates, our Privacy-Aware (PA) model prioritizes semantic generalization.  Furthermore, unlike the existing methods that transmit the coordinates data to third party API \cite{zhang2026narrativesense, zhang2024aware}, our transformation pipeline leverages a self-hosted OpenStreetMap (OSM) engine and a Large Language Model (LLM) to transform raw GPS data directly into behaviorally meaningful, human-understandable categories (e.g., home, school, recreation, work) in real-time. 

By implementing this framework, we offer a pathway for researchers to safely collect and transform location data and provide potential method for release behavioral datasets while empowering the broader community to build Explainable AI (XAI) models. In this work, our core contributions are:

\begin{enumerate}
    \item A Location Data Transformation Method for Data Collection and Public Data Sharing: We propose the PA framework as a standardized preprocessing step that allows data stewards to completely destroy sensitive geographic identifiers while retaining the rich behavioral utility necessary for high-performance stress recognition.
    \item Mitigating the Privacy-Explainability-Utility Trade-off: We empirically demonstrate that semantic generalization of GPS trajectories provides robust resistance against re-identification attacks (reducing top-1 adversary accuracy to 15\%) while preserving the human-readable, clinically actionable explainability (via SHAP) required for user and clinician trust.
    \item Flexible and Private Semantic Pipeline: We introduce a privacy-by-design architecture utilizing a self-hosted OSM engine server and LLM-bootstrapping for location feature engineering, ensuring zero sensitive coordinates data leakage to third-party commercial APIs.
\end{enumerate}

\section{Background and Related Work} \label{rw}

\begin{figure}[!ht]
    
    \includegraphics[width=\linewidth]{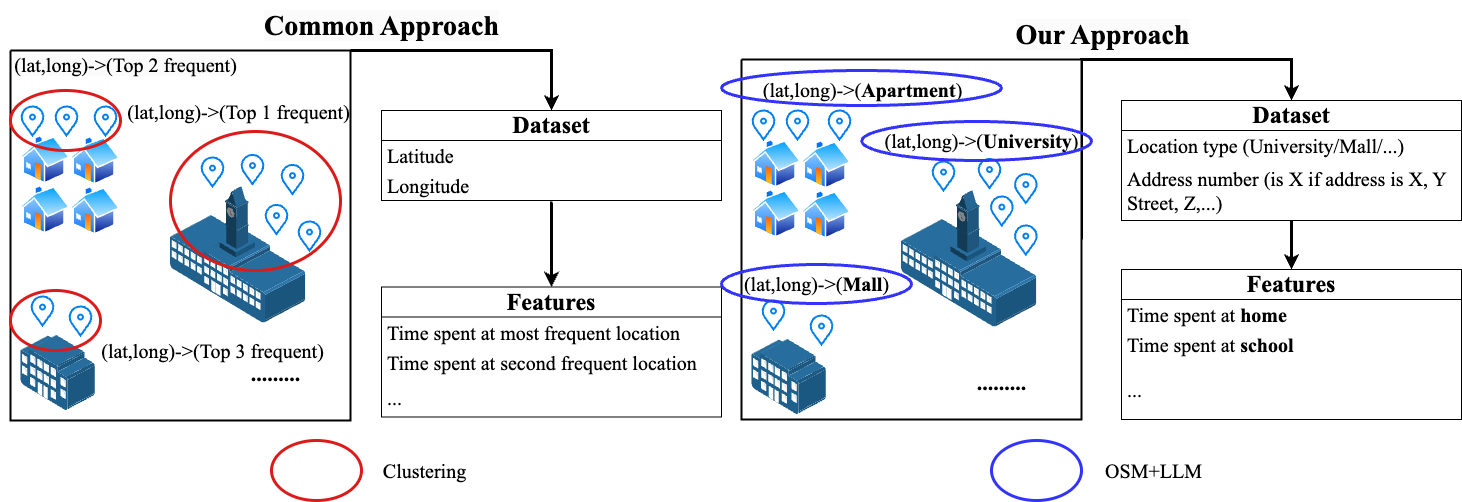}
    \caption{The comparison of the current common GPS feature extraction method and our method. While GPS signals must be stored in a dataset, which poses privacy threats from exposure of locations, to examine the top frequent locations in the common method, our method showcases a GPS signal transformation technique, by self-hosted OSM and LLM for instant transformation, to secure the privacy of the user's location and prevent information loss from a less frequently visited address.}
    \label{fig:enter-label}
\end{figure}
Passive mental health monitoring has been developed as a method to enhance human quality of life. This is often achieved by leveraging data from ubiquitous technologies like smartphones and wearable devices. For instance, smartphones, with their array of sensors, can provide valuable insights into a user's behavior, including movement patterns, conversation frequency, and sleep quality, which can be correlated with mental well-being \cite{khoo2024machine}. 
\subsection{Feature Extraction and Modeling for Mental Health Recognition}

\subsubsection{Critical Gaps in Traditional Behavioral Sensing and Location Modeling}
The StudentLife dataset \cite{student_life}, a comprehensive collection of smartphone-derived data, provides a valuable foundation for examining student behavior and its impact on academic outcomes. Wang et al. leveraged this dataset to explore the correlation between daily living patterns and student learning, extending beyond traditional academic metrics. Upon the StudentLife research, many studies \cite{student_life4,student_life2,student_life3} leverage various mobile sensing data for their application, such as conversation, activity, etc., for mental health assessments. This research used RAPIDS \cite{vega2021reproducible}, a framework built for feature extraction from all resources from mobile sensing data, such as acceleration, conversation, location, etc, as the backbone for their application.  However, this approach raises serious concerns about the effective utilization of mobile sensing data, privacy threats, and transparency for mental health assessment applications.

Furthermore, most prior works in mobile sensing for mental health recognition used density-based spatial clustering of applications with noise (DBSCAN) \cite{10.5555/3001460.3001507} for localizing frequent locations, but present notable limitations. DBSCAN, while effective in identifying clusters of prolonged presence, inherently discards data related to less frequent activities, such as recreational outings or travel. This results in a substantial loss of potentially valuable information about less time-consuming and random activities as recreation or traveling. More critically, this method also assumes the longest cumulative stay is home, which might not reflect the right function of location, causing a misinterpretation in an explainable system.

\subsubsection{Advancements and Vulnerabilities in Deep Learning and LLM-based Sensing}
The state-of-the-art, Deep Spatio-Temporal Graph Neural Network (STGNN) approach has demonstrated strength in handling GPS data. For instance,  \citet{harit2024monitoring} showed that using only raw GPS data, an STGNN improved stress classification by 3\%, increasing the accuracy and F1 score from the 58\% and 55\% achieved by traditional GCN methods. However, such deep learning approaches present challenges, including a lack of model transparency and, more importantly, significant privacy concerns due to their reliance on precise location coordinates that make it unsuitable for deployment.

Recent advancements have attempted to leverage Large Language Models to extract semantic behavioral insights directly from passive sensing data. For instance, Zhang et al. \cite{zhang2026narrativesense} recently proposed NarrativeSense, which utilizes the AWARE Narrator tool \cite{zhang2024aware} to predict affective states. While this approach successfully captures rich semantic context, it relies heavily on transmitting sensitive raw location coordinates to third-party commercial services (e.g., the Google Maps API) for reverse-geocoding. Furthermore, by utilizing unprotected location identities and raw timestamps as features, this methodology exposes users to severe re-identification vulnerabilities. As we quantitatively demonstrate in our GeoLife ablation study (Section \ref{ablation_study}), exposing raw location types allows an adversary to achieve near-perfect re-identification accuracy (over 96\%). This highlights a critical flaw in current semantic mapping tools: achieving contextual utility by sacrificing architectural privacy.

\subsection{Privacy-preserving Methods and Limitations}
\subsubsection{Traditional Data Transformation}
While robust privacy-preserving methods are well-established for data types like speech \cite{abdullah2018sensing}, text \cite{sousa2023keep}, and facial images \cite{chen2018vgan}, their application to location data presents unique challenges. Formal techniques, such as differential privacy \cite{xiao2015protecting} for location-based applications, provide mathematical guarantees by altering the original data through noise injection or mathematical transformation, which is often associated with a high cost to data utility. In the context of behavioral modeling, such an approach can alter the data so significantly that it obscures the subtle nuances of a person's location experience, which in turn affects the \textit{explainability} and \textit{user trust} of the application.

Another important method for protecting location data is obfuscation (or spatial cloaking) \cite{yin2015re, baten2023predicting}, where coordinates are intentionally made less precise to enhance user privacy. However, this approach still involves either storing users’ sensitive location data, which poses a significant privacy risk, or using third-party API (for example, Foursquare \cite{morshed2019prediction}, Google Maps API \cite{zhang2026narrativesense,zhang2024aware}) to perform data transformation, which poses a threat by \textit{sharing data} to third-party services. 
\subsubsection{Federated Learning Method}
Federated Learning \cite{mcmahan2017communication}, a decentralized approach to machine learning deployment, is another noteworthy technique for enhancing user privacy. In this framework, edge devices (e.g., smartphones, edge devices, etc.) participate in model training alongside a central server. The devices locally train the model on their data and then send only the updated weights, subject to predefined constraints, to the server, which aggregates them to update the global model. The updated global model is then distributed back to the participating devices. This process ensures that users’ raw data never leaves their devices, thereby offering strong privacy protection.

Nevertheless, recent studies have shown that attackers can exploit vulnerabilities in this setup. For instance, they may reverse-engineer model weights to reconstruct sensitive data (reconstruction attacks) \cite{zhu2019deep,geiping2020inverting}, or poison the model during training (model poisoning attacks) \cite{yazdinejad2024robust}. These threats highlight persistent privacy and model safety risks for users despite the advantages of federated learning.

These limitations of existing methods highlight the need for a more granular framework specifically designed for the complexities of location information. \citet{liu2018location} provides such a framework by identifying three key aspects of location privacy: identity, spatial, and temporal. Adopting this multi-faceted view allows for the development of more targeted privacy-enhancing solutions that can better balance data protection with the utility required for specific applications.

Guided by this structured understanding, our work develops a location encoding method (PA) that directly addresses the challenges outlined by \citet{liu2018location}'s framework. Instead of applying a uniform privacy mechanism, our approach, which is detailed in Subsection \ref{method}, is specifically designed to mitigate spatial and temporal privacy risks. By transforming raw coordinates into meaningful categories and analyzing aggregated time-based features, we preserve the essential behavioral signals for our model while enhancing the privacy of the user's location data.

\subsection{Current Location Data Sharing Limitations}

While continuous location tracking yields invaluable insights into human mobility and mental well-being, raw GPS trajectories are inherently too sensitive to share externally. To comply with privacy regulations and protect participant identities, researchers releasing open-source longitudinal datasets are currently forced to mathematically obfuscate or aggregate this data prior to publication \cite{student_life2,zhang2025causalcff}. Traditionally, this transformation process strips away all geographic context, categorizing mobility into four primary abstract feature sets:

\begin{enumerate}
    \item Clustering-based features: This approach represents mobility as the total duration spent at a user's "Top $k$" most frequented spatial clusters. In this paradigm, "Cluster 1" is often blindly assumed to be the user's home, which frequently causes critical misalignments with the actual function of the location.
    \item Entropy-based features: These metrics quantify the mathematical unpredictability or chaotic nature of a user's daily routine, without capturing the actual environmental context of their whereabouts.
    \item Kinematic and Movement features: This set contains purely physical metrics, such as total distance traveled, variance in speed, or location transition frequency.
    \item Noise Injection: This technique involves introducing random perturbations into the dataset’s coordinates and/or timestamps to preserve data privacy. However, a significant drawback is that it often degrades data interpretability, leading to an unexplainable feature set.
\end{enumerate}

The fundamental limitation of these existing abstraction methods is the complete destruction of semantic meaning. When datasets are published using these abstract metrics solely, downstream researchers are severely handicapped. Features such as "duration at Top k location" or "high location entropy" cannot be intuitively mapped to specific human behaviors or psychological states. Consequently, researchers utilizing these public datasets are forced to rely on opaque, black-box deep learning models to find hidden correlations, rendering the resulting mental health inferences clinically unexplainable to both patients and healthcare providers. 

Our method presents a paradigm-shifting alternative for data transformation prior to public release. Instead of obfuscating raw coordinates into meaningless geometric clusters, our framework transforms them directly into human-understandable, functional behavioral categories (e.g., time spent in recreation, at work, or at home). By adopting this semantic transformation as a standard for data sharing, data stewards can completely destroy sensitive geographic identifiers while fully preserving the behavioral utility of the dataset. Ultimately, this empowers the broader ubiquitous computing community to develop highly accurate, privacy-preserving, and, crucially, explainable models that provide actionable, human-readable insights.

\section{Methodology}
\subsection{Dataset}
As modern datasets often pre-cluster the location data before deployment, the StudentLife dataset \cite{student_life} was chosen as a deliberate benchmark for feature mining and insight analysis in this study. 
This dataset, gathered in 2014, utilized an Android app to consistently track various data streams, including GPS, conversation, activity recognition, etc., alongside monitoring the mental health status, deadlines, etc., of the participants. The dataset comprises information from 48 students over a 10-week academic term at Dartmouth College.
In addition, notwithstanding the age of the data, the StudentLife dataset continues to serve as a canonical benchmark in computational behavior recognition \cite{harit2024monitoring,luo2024dynamic}.

This research specifically focuses on utilizing subsets of the dataset, namely the GPS (recorded in 20-minute intervals) and academic information for analysis and exploration.
% time, provider, type of network, latitude, and longitude.

% \begin{table}[!ht]
% \caption{The table of GPS dataset features being used}
% \begin{center}
% \begin{tabular}{c  c}
% \hline
% \textbf{Feature name}&\textbf{Definition}\\
% \hline
% \textbf{Time} &  Unix Timestamp  \\

% \textbf{Latitude} &The latitude of GPS signal\\

% \textbf{Longitude} & The longitude of GPS signal \\
% \hline
% \end{tabular}
% \label{tab1}
% \end{center}
% \end{table}

In this work, time, longitude, and latitude from the GPS data stream from StudentLife research \cite{student_life} were utilized as the source for GPS data. Apart from that, we also used 3 sets of StudentLife academic-based data streams, shown below, to enrich the feature set:
\begin{itemize}
  \item Class stream: contains the classes of each volunteer in the examined semester.
  \item Class info stream: contains the information of each class: class name, location, and period.
  % (consisting of weekday as decimal number and start time and end time of the class period).
  \item Deadline stream: contains the deadlines of each day for every volunteer. 
  % From the deadline, we could understand more about academic pressure from student life.
\end{itemize}

Finally, we used the EMA stress response dataset for labeling the stress status of the students. 
In this dataset, the stress label is collected at 5 levels: a little stressed, definitely stressed, stressed out, feeling good, and feeling great, and labeled from 1 to 5, respectively. 
However, this label does not show a linear progression between the stress levels and the label. Hence, we employ label transformation, which transforms labels from 1 to 5 to represent feeling great, feeling good, a little stressed, definitely stressed, and stressed out.

Following that, for some days, receive multiple stress reports, the final stress report will be the median of the report values. After that, a binary stress level was employed by assigning a stress level of less than 3 to no stress (0) and stress otherwise (as suggested in \cite{harit2024monitoring,bonafonte2023analyzing}). Ultimately, we received 2780 daily data records for re-identification attack and 1131 daily stress report records for stress recognition from the StudentLife dataset.

Additionally, while StudentLife is our primary dataset for evaluating stress recognition utility, we separately utilize a subset of the GeoLife dataset \cite{zheng2009mining,zheng2008understanding,zheng2010geolife} (detailed in Section \ref{ablation_study}) as a standalone ablation study to confirm the scalability and robust privacy-preserving capabilities of our encoding framework.
\subsection{Location Encoding Method} \label{LEM}
The influence of environmental context on human emotional states is well-documented, with studies indicating that exposure to natural environments, such as parks and green spaces, often elicits positive affective responses, while confined or stressful environments, like workplaces, can contribute to heightened stress levels \cite{olszewska2022features,koivisto2023mental}. This phenomenon is attributed, in part, to the activation of the autonomic nervous system in response to perceived environmental cues.

In this research, to quantify the relationship between student location patterns and their potential impact on well-being, our method is designed to transform raw GPS coordinates into semantic features in a privacy-aware, two-step process:

First, raw coordinates are mapped to address types and numbers. The implementation of this step involves a self-hosted reverse-geocoding engine using a pre-downloaded OpenStreetMap (OSM) \cite{OpenStreetMap} database. The OSM engine (Nominatim) was selected for its lightweight, easy-to-deploy, and memory-efficient architecture\footnote{See https://nominatim.org/release-docs/latest/admin/Installation/ for installation and self-hosting documents}. Requiring less than 3 GB of combined storage for the Dartmouth/Hanover and Beijing regional maps, it enables rapid, localized reverse-geocoding on standard self-hosted servers without the need for specialized hardware. This architecture ensures no identifiable GPS data is transmitted to a third party server. Then the coordinates are sent to a self-hosted server for reverse geocoding. As a result, we collected 103 distinct location types and identities. This location identity, which is used to extract the number of locations visited, is made by concatenating the address number and address type.

Second, these 103 granular location types are classified into key life-related categories. LLMs, which have been shown to be powerful and highly accurate (up to 92\%) for recognizing location functions \cite{liu2024semantic}, offer a valuable capability for this categorization, especially in real-world deployments where the system encounters previously unseen edge cases. Our method implements this as a static, pre-defined map (e.g., dormitory $ \rightarrow $ home) that runs locally. This ensures minimal external API calls, triggering a dynamic request only when an unmapped location type is encountered, thus preserving system privacy and minimizing latency \cite{baten2023predicting}.

To develop this foundational map, we employed the public Gemini 2.5 Flash LLM \cite{geminiteam2024geminifamilyhighlycapable} as a bootstrapping tool, selected for its clear privacy policy. We utilized the following zero-shot prompt: "Please classify location\_type into one of the following groups: home, school, shop, workplace, recreation, travel, and others. Respond with only the group name." This prompt-driven approach not only provides rapid location categorization but also introduces immense architectural flexibility. Rather than retraining classification algorithms or manually re-labeling thousands of OpenStreetMap tags, future researchers can seamlessly adapt the system to target specific behavioral domains (e.g., isolating food venues or medical facilities) simply by modifying the zero-shot prompt. For the StudentLife benchmark in this study, 103 initial LLM calls were performed to construct the localized dictionary, efficiently converting the raw location types into semantic, function-based features.

To validate the LLM's output for StudentLife dataset, we created a ground truth by manually labeling all 103 types. The LLM's classification achieved 91.3\% zero-shot accuracy against this ground truth, ensuring the reliability and ethical grounding of the static map used in our research. 
In real system deployment, dynamic call to LLM (both online or offline), which only transmits the abstract string of the unseen OSM location type, will be utilized to update the location conversion database to handle the edge case, unforeseen scenario in real-life seetings.

\subsection{Feature Extraction}

From the GPS and academic data, a total of 54 features were extracted in this work. 
According to the extraction method and behavior insight, the features were divided into 3 groups: Location-based features, Academic-based features, Time-based features

\begin{table}[!ht]
    
        \caption{The table of the feature types and quantities for each evaluation setting. The STGNN baseline is from Harit et al.'s work \cite{harit2024monitoring}. $(^*$) indicates that no extracted features were used, as the deep learning method follows strictly Harit et al.'s work.}
    \begin{tabular}{cccccccc}
    \hline
     \textbf{Data stream}& \multicolumn{2} {c} {\textbf{Feature type}}  &\textbf{Quantities}  &\textbf{AF}  & \textbf{PA}& \textbf{STGNN GPS}& \textbf{STGNN All}\\
       \hline
     \multirow{3}{*}{\textbf{GPS}}  & \multirow{2}{*}{\textbf{Location}}  & \textbf{Function-based} & 38&  \checkmark & \checkmark&  \checkmark & \checkmark\\
 & & \textbf{Address-based} & 3& \checkmark &&  \checkmark & \checkmark\\
&\multicolumn{2} {c} {\textbf{Time-based}}&  2 &\checkmark &\checkmark &  \checkmark & \checkmark\\
\hdashline
    \textbf{Class and Deadline}   & \multicolumn{2} {c} {\textbf{Academic-based}}&  11 & \checkmark& &   & \checkmark\\
    \hdashline
    \textbf{Activity, Wifi}&&&&&&&\multirow{2}{*}{\checkmark}\\
    \textbf{and Phone log}\\
         
         \hline
        & \multicolumn{2} {c} {\textbf{Number of Features}}&54&54&40&N/A$^*$&N/A$^*$\\
         \hline
    \end{tabular}

    \label{AFPA}
\end{table}

\subsubsection{Location-based Features} \label{mob_feat}

As the environment plays a pivotal role in human feeling, we then calculate the total time spent and standard deviation by each participant in each location function (see Section \ref{LEM}), during defined daytime (6:00 AM to 6:00 PM), nighttime (6:00 PM to 6:00 AM), and only total time spent in the whole day (24 hours) period for location-function-based features. To quantify time spent outside of primary academic and domestic environments, we compute the cumulative time spent in all categories, excluding 'home' and 'school'. To assess the balance between academic and personal life, we calculate the difference between daytime and nighttime durations spent at home and school, as expressed in equation \eqref{daytimevsnighttime}:
\begin{equation}
feature\_time_{d\_vs\_n}=feature\_time_{d}-feature\_time_{n}
\label{daytimevsnighttime}
\end{equation}

where feature represents either 'home' or 'school' and d and n represent day and night, respectively. Ultimately, 38 location-function-based features ($2\times 7 +2\times7+1\times7+1+2$) were extracted.

To quantify mobility patterns, 3 location-address-based features were extracted. Firstly, we collected location address numbers to determine the number of unique locations visited via the total number of different address identities. Furthermore, we defined "repetition" as the number of instances a participant visits the same address identity (e.g., 10 Apartment) within a specified timeframe. Mobility activity repetition metrics will be calculated on both a daily and weekly basis.

\subsubsection{Academic-based Features}

Academic stressors, such as demanding class schedules and impending deadlines, are prevalent in student life, often leading to heightened levels of exhaustion and stress. Therefore, this study aims to quantify the relationship between class schedules, attendance patterns, deadline proximity, and student stress.

 In this research, we first extracted class schedule data from class information and the class dataset. Attendance rates were then calculated by comparing scheduled class times with time spent in academic locations, as determined by GPS data (refer to Section \ref{mob_feat} for details on location data processing). An attendance rate below 0.7 was defined as an indicator of potential class absenteeism. We then examined instances of potential class absenteeism across five temporal windows - the day of, the day after, two days after, three days after, and seven days after the scheduled class - to assess the temporal impact of absenteeism on student stress.

Furthermore, we extracted deadline information from course syllabi to quantify the proximity of upcoming assignments. We assessed the status of deadlines on the day of data collection and then examined the three days preceding the deadline (three, two, and one day prior) to analyze the escalating impact of deadline proximity on student stress.

\subsubsection{Time-based Features}

Time is an omnipresent factor in human life, influencing behavior and thought patterns in profound ways. For example, individuals often experience a sense of anticipation as Friday or holiday weeks approach \cite{stone2012day}, while feelings of fatigue and resignation can become more pronounced as Monday comes to a close \cite{areni2008memories}.\\
Recognizing these temporal variations, we have incorporated two key features into our analysis: the "date in week" and the "week index." These features enable us to capture the dynamic role of time in shaping student mental stress behavior across an academic semester. By accounting for both the specific day of the week and the position within the semester, our approach allows for a nuanced understanding of how academic pressures affect mental well-being during periods such as midterms, finals, and weekends.

\subsection{Threat Modeling}\label{TM}
% adding soon

In ubiquitous computing systems, the risk of data being exploited by malicious actors is a primary concern for user privacy. This section presents a simulation to quantify the privacy-preservation guarantees of our encoding strategy against a re-identification attack.

In this work, we define a threat model where an attacker has gained access to the feature database (see Figure \ref{fig:enter-label}). The attacker's objective is to re-identify specific users from the supposedly anonymized data. We simulate three scenarios based on the attacker's prior knowledge, corresponding to the amount of data they possess:
\begin{enumerate}
    \item Rich Knowledge: The attacker has a large dataset of known user behavior (80\% train, 20\% test).

\item Moderate Knowledge: The attacker has a balanced dataset (50\% train, 50\% test).

\item Limited Knowledge: The attacker has very little prior data (20\% train, 80\% test). This scenario reflect upper bound of a realistic attack attempt on trying to identify a person \cite{biggio2013evasion,tournier2022expanding}.
\end{enumerate}

To evaluate the risk, we compare the re-identification accuracy across three feature sets:

\begin{enumerate}
    \item Privacy-Aware (PA): Our proposed privacy-enhanced model, which removes the high-risk repetition-based and academic-based features.

\item All-Features (AF): The full feature set, which includes the daily and weekly location repetition features identified as a privacy risk.

\item Raw GPS Baseline: A set of five simple statistical features (mean, max, min, standard deviation, and inter-quartile range) extracted directly from the volunteers' daily latitude and longitude coordinates.

\end{enumerate}

For each of the three "knowledge" scenarios, we trained an XGBoost classifier to predict the user's ID. The model's accuracy on the test set represents the success rate of the attack. A lower accuracy score indicates a more robust defense against re-identification and, therefore, a stronger privacy guarantee.

\subsection{Stress Classification Model}\label{CLM}

In this study, we employed two ensemble learning models, Random Forest and XGBoost with random search, to predict daily stress levels.

\begin{itemize}
  \item Random Forest (RF) is a learning model that, according to the Gini impurity factor, combines decision trees, and the decision is based on the Gini coefficient. The result of the RF model is the average of the results from the decision trees within it. It aggregates results from multiple decision trees to create a model with low bias and variance \cite{rf}. In this study, base RF was tuned with \textit{n\_estimators=70}.
  
  \item XGBoost (XGB) is an optimized distributed gradient boosting library designed to be highly efficient, flexible, and portable\cite{xgb}. It implements machine learning algorithms under the Gradient Boosting framework. XGBoost provides a parallel tree boosting that solves many data science problems in a fast and accurate way. This model was set with \textit{eta=0.2} and \textit{subsample=0.5}.

\end{itemize}
To evaluate the efficacy of our proposed features, we benchmark the performance of our RF and XGB models against the deep learning baselines reported by Harit et al. \cite{harit2024monitoring} on the StudentLife dataset, specifically their LSTM, STGNN GPS, and STGNN All configurations.
\subsection{Evaluation Settings}\label{method}
After feature extraction, two feature sets were examined for students' stress recognition, including PA and AF (see table \ref{AFPA}).

These feature sets were evaluated by accuracy (Acc) and F1 score (F1) in three conditions: 75:25 random split and LOSO validation. In each condition, we acknowledge an imbalance between the positive and negative stress classes. Therefore, after training and validation separation, the data will be balanced via the Edited Nearest Neighbour Synthetic Minority Over-sampling Technique (SMOTEENN) \cite{10.1145/3055635.3056643} for a 75:25 random split and  Synthetic Minority Over-sampling Technique (SMOTE) \cite{10.5555/1622407.1622416} for LOSO validation. Furthermore, to ensure reliable LOSO validation, participants with fewer than 10 days of recorded data were excluded from the test set.
\section{Results and Analysis} \label{mainresult}
\subsection{Privacy Analysis}

\subsubsection{Re-identification Attack Accuracy}

\begin{table}[!ht]

    \caption{The re-identification attacks accuracy (top 1 accuracy (top 5 accuracy)) in 3 scenario (in percent). }
    \begin{tabular}{cccc}
    \hline
        & \textbf{AF} & \textbf{PA} & \textbf{Raw GPS}\\
        \hline
        \textbf{Rich Knowledge} & 41 (68)  & 25 (52) & 73 (92)\\
       \textbf{Moderate Knowledge}  & 30 (62)  & 22 (46) & 69 (89)\\
       \textbf{Limited Knowledge}  & 19 (43) &  15 (38)& 57 (82)\\
       
    \hline
    \end{tabular}
    
    \label{tab:reid}
\end{table}

\begin{figure}[ht!]
    \centering
    \includegraphics[width=\linewidth]{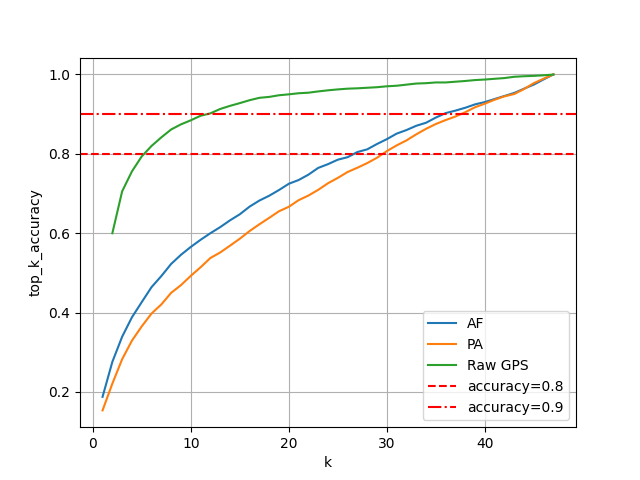}
    \caption{The top-k-accuracy of the re-identification attack in the \textit{Limited Knowledge} scenario. The Raw GPS method \cite{harit2024monitoring} (green) reaches high accuracy at a low $k$, while our PA (orange) and AF (blue) methods require a significantly larger $k$ to achieve similar accuracy.}
    \label{fig:topKacc_comp} % Changed label to fig: for clarity
\end{figure}
The re-identification simulation attack demonstrates the significant privacy-preserving capabilities inherent in our proposed encoding techniques. Both of our methods, PA and AF, yield substantially lower re-identification accuracy metrics when contrasted with the unprotected Raw GPS feature baseline. The efficacy of this mitigation is particularly pronounced in our PA approach. 

Examining the Top-1 accuracy (as detailed in Table \ref{tab:reid}), the PA method consistently reduces the re-identification success rate to approximately one-half to one-third of that achieved using Raw GPS data. This stark decrease provides compelling evidence for the method's ability to thwart such re-identification attempts.

A similar and equally important trend is observed when analyzing the Top-5 re-identification accuracy. The PA model not only demonstrates a substantial accuracy reduction compared to the Raw GPS baseline (e.g., 38\% vs. 82\% in the Limited Knowledge) but also consistently outperforms the AF method, offering a further 5-8\% absolute reduction in accuracy. This highlights its superior capability in obfuscating user identity even when the attacker broadens their search criteria.

This privacy-enhancing effect is further visualized in the top-k accuracy plot, presented in Figure \ref{fig:topKacc_comp}. This figure illustrates the number of candidates ($k$) an attacker must consider to achieve a given re-identification accuracy. For an attacker using the Raw GPS data, moderate (80\%+) and high (90\%+) re-identification accuracy can be achieved with a very small candidate set of $k \approx 5$ and $k \approx 10$, respectively. This indicates a severe privacy leak, as an individual can be easily isolated.

In stark contrast, our proposed methods fundamentally alter this dynamic. To achieve the same 80\% accuracy threshold, our approaches require the attacker to expand their search to $k \approx 25 $. To reach 90\% accuracy, this figure increases to $k \approx 35$. This effectively means our encoding techniques make the re-identification task 3.5 to 5 times more difficult, significantly increasing the ambiguity and computational burden for any adversary. By forcing the attacker to consider a much larger anonymity set, our methods provide a robust and practical defense against re-identification.

\subsubsection{Mutual Information}
\begin{table}[]
 \caption{The table of the top 10 Mutual Information values between the Feature and the volunteer identity.}
 \begin{tabular}{cccc}
 \hline
\textbf{Feature} & \textbf{Mutual Information} & \textbf{AF} & \textbf{PA} \\
\hline
 class\_schedule & 0.695339 & \checkmark & \\
 home\_time & 0.357262 & \checkmark & \checkmark \\
 home\_time\_noon & 0.325728 & \checkmark & \checkmark \\
home\_time\_day\_vs\_night & 0.319390 & \checkmark & \checkmark \\
school\_time\_day\_vs\_night & 0.298065 & \checkmark & \checkmark \\
 School\_and\_home\_off\_time & 0.281064 & \checkmark & \checkmark \\
 home\_time\_daytime & 0.264064 & \checkmark & \checkmark \\
 school\_time\_noon & 0.262472 & \checkmark & \checkmark \\
 number\_of\_location\_visited & 0.252563 & \checkmark & \\
 recreational\_activities\_time & 0.245008 & \checkmark & \checkmark \\
 \hline
\end{tabular}

 \label{MI}
\end{table}

To better understand exactly which behavioral features drive the re-identification vulnerabilities observed in the previous section, we calculated the Mutual Information (MI) between individual features and the volunteers' identities (see Table \ref{MI}).

 Overall, the PA model has eliminated the strongest feature, class\_schedule, which has the highest MI value. This indicates that the PA model successfully removes the feature most correlated with user identity, thereby enhancing privacy. Furthermore, another high-MI feature, number\_of\_location\_visited, is also excluded from the PA set.

This analysis highlights the core privacy-utility trade-off. The PA model retains eight of the top ten MI features, including various metrics for 'home\_time' and 'school\_time'. These features, while still containing some identity-related information (MI values from 0.245 to 0.357), are more general and less unique than a specific schedule. By keeping these moderately-correlated features, the model preserves its predictive utility for the stress recognition task while significantly enhancing user privacy by removing the most egregious identifying features.

\subsection{Stress Recognition}
\subsubsection{Random Splitting}

\begin{table}[!ht]
    
    \caption{The table of classification performance comparison in the random splitting method to the baseline (in \%).}
    \begin{tabular}{ccccccccc}
    \hline
    
        \multirow{3}{*}{\textbf{Model}} &\multicolumn{3}{c}{\textbf{Baseline}} &\multicolumn{2}{c}{\textbf{PA}}&\multicolumn{2}{c}{\textbf{AF}}\\
        &\textbf{  STGNN}& \multirow{2}{*}{\textbf{LSTM}}& \textbf{STGNN} &\multirow{2}{*}{\textbf{XGB}}&\multirow{2}{*}{\textbf{RF}} &\multirow{2}{*}{\textbf{RF  }}&\multirow{2}{*} {\textbf{XGB}}\\
        &\textbf{GPS}&&\textbf{All}&&\\
        \hline
        \textbf{Acc}&  61&56 &  \textbf{67} & 65& \textbf{67} &  \textbf{68}&65  \\
     
        \textbf{F1}&  58  &  54 &  \textbf{65}&   62& \textbf{64}&  \textbf{65}  &  60  \\
        % \textbf{AUC-ROC}& N/A  &  N/A   &\textbf{60}  & 56  & 58  & \textbf{60}\\
        % afrf 68 acc 65 f1
         \hline
    \end{tabular}
    \label{rs}
    
\end{table}
% Table \ref{rs} summarizes the classification performance, highlighting the substantial recognition capability of the proposed methods.  Significantly, all presented approaches exhibited superior F1 scores in comparison to the STGNN baseline \cite{harit2024monitoring} in GPS scenarios.  The PA\_RF model, identified as the optimal configuration, attained a compelling F1 score and accuracy of 64\%, reflecting a 6\% enhancement in F1 score and accuracy, respectively, relative to the STGNN GPS baseline \cite{harit2024monitoring} and  10\% F1 compared to LSTM.
Table \ref{rs} summarizes the classification performance of our proposed frameworks, PA and AF, against the key baseline models.

A critical observation is that our proposed approaches, when paired with appropriate classifiers, consistently outperform the standard Deep Learning, LSTM and STGNN GPS \cite{harit2024monitoring}, baselines. The STGNN GPS baseline achieves an F1-score of 58\%, while the LSTM model performs at 54\%. In significant contrast, all variants of our PA and AF models exhibited superior F1-scores, demonstrating their effectiveness.

The AF\_RF model achieves the highest overall performance in the evaluation, with a peak accuracy of 68\% and a top-tier F1-score of 65\%. This result is particularly noteworthy as it not only surpasses the specialized STGNN GPS model but also matches or even exceeds the performance of the STGNN All baseline (67\% Acc, 65\% F1), which has the advantage of using all available features without any privacy considerations.

Furthermore, our PA\_RF model, which we identify as the optimal configuration for balancing utility with the privacy-preservation goals demonstrated earlier, attains a highly compelling performance. It achieves an accuracy of 67\% and an F1-score of 64\%. This reflects a substantial 6\% absolute enhancement in F1-score (64\% vs. 58\%) and in accuracy (67\% vs. 61\%) relative to the STGNN GPS baseline \cite{harit2024monitoring}. The improvement is even more pronounced when compared to the LSTM baseline, showing a 10\% absolute lead in F1-score (64\% vs. 54\%).

Crucially, the utility of our privacy-aware PA\_RF model (67\% Acc, 64\% F1) is statistically on par with the non-private, feature-rich STGNN All baseline (67\% Acc, 65\% F1). This outcome strongly validates our approach, proving that our method can successfully mitigate re-identification risks while simultaneously preserving a high degree of data utility necessary for complex downstream classification tasks.
% Additionally, for privacy enhancing, only a minimal 2\% drop in F1-score as a trade back with .

% \begin{table}[!ht]
    
%     \caption{The table of classification performance comparison in the random splitting method between scenarios (in \%).}
%     \begin{tabular}{ccccccc}
%     \hline
    
%         \multirow{3}{*}{\textbf{Model}} &\multicolumn{3}{c}{\textbf{AF}} &\multicolumn{2}{c}{\textbf{PA}}\\
%         &\textbf{RF  }& \textbf{XGB}& \textbf{DC} &\textbf{XGB}&\textbf{RF} \\
%         \hline
%         \textbf{Acc}&  \textbf{68}&65 &  31 & 65& 67  \\
     
%         \textbf{F1}&  \textbf{65}  &  60 &  24&   62& 64 \\
%         % \textbf{AUC-ROC}& N/A  &  N/A   &\textbf{60}  & 56  & 58  & \textbf{60}\\
%         % afrf 68 acc 65 f1
%          \hline
%     \end{tabular}
%     \label{sc}
    
%     \label{tab:my_label}
% \end{table}

When comparing the AF and PA settings (refer to Table \ref{rs}), there is a minimal trade-off in performance. Specifically, the AF model using RF (AF\_RF) achieved an F1-score of 65\%, which is just 1\% better than the PA setting with RF.
\subsubsection{Leave-one-subject-out (LOSO)}
\begin{table}[!ht]
    
    \caption{The table of classification performance in  LOSO cross-validation method (in \%)}
    \begin{tabular}{ccccc}
    \hline
    
        \multirow{2}{*}{\textbf{Model}} &\multicolumn{2}{c}{\textbf{PA}} &\multicolumn{2}{c}{\textbf{AF}}\\
        &  \textbf{RF} &\textbf{XGB}& \textbf{RF}  &\textbf{XGB}\ \\
        \hline
        \textbf{Acc}& 62\textpm14 & 61\textpm18  & 62\textpm15 & 60\textpm14 \\
        \textbf{F1} & 70\textpm16 & 70\textpm15 &  71\textpm15&    69\textpm16 \\
         \hline
    \end{tabular}
    \label{LOSOCV}
    
    \label{tab:my_label}
\end{table}
Table \ref{LOSOCV} presents a comprehensive comparison of the classification performance for our proposed models, evaluated using the highly rigorous Leave-One-Subject-Out (LOSO) cross-validation method. This validation paradigm represents a stringent test of a model's ability to generalize, as it must make predictions for an entirely unseen subject after being trained on all other subjects. The data from this evaluation emphatically show a comparable level of performance across the four main classifier configurations (PA RF, PA XGB, AF RF, and AF XGB).

This consistency is a critical finding, as it underscores the robustness of our feature-encoding and classification strategy. The fact that the models perform consistently even when subjected to this challenging validation scheme suggests that they are learning generalizable patterns from the data rather than overfitting to subject-specific biases or idiosyncrasies. This significantly enhances our confidence in the broader applicability and real-world utility of the findings.

Specifically examining the accuracy metric, all four primary models exhibit remarkably similar mean performance, with values tightly clustered in the 60\% to 62\% range (62\% for PA RF, 61\% for PA XGB, 62\% for AF RF, and 60\% for AF XGB). Despite these slight numerical variations in the mean values, the associated standard deviations are substantial, spanning approximately $\pm$14\% to $\pm$18\%. This high variance is characteristic of LOSO validation, where performance can fluctuate depending on the specific subject left out. The extensive overlap in these standard deviation ranges definitively indicates a lack of any statistically significant differences in accuracy across these four models.

This pattern of homogeneity is further reinforced when analyzing the F1 scores. Here, the models again show strong, consistent results, with mean scores residing in a narrow band around 70\% (70\% for PA RF, 70\% for PA XGB, 71\% for AF RF, and 69\% for AF XGB). As with accuracy, the associated standard deviations are large (around $\pm$15\% to $\pm$16\%), confirming that these minor differences in mean F1 scores are not statistically meaningful.

\subsubsection{Feature Analysis and Explainability}
\begin{table}[!ht]
    \centering
    \caption{The top-10 significant features ranked by F-test p-values and their corresponding r-values.}
    \begin{tabular}{ccc}
    \hline
      \textbf{Features}   & \textbf{p-value} &\textbf{r-value} \\
      \hline
        Week & 2.8e-7 & 0.15\\
       2\_days\_after\_skip\_class  & 7.6e-5 & 0.12\\
        working\_time  & 1.8e-3 & -0.09\\
        travel\_time & 2.8e-3 & 0.09\\
        deadline  & 5.0e-3 & 0.08\\
        recreational\_activities\_time  & 5.0e-3 & -0.08\\
        1\_day\_to\_DL  &6.6e-3  & 0.08\\
        number\_of\_location\_visited  &2.2e-2  & -0.07\\
        3\_day\_to\_DL & 3.8e-2 & 0.06\\
        week\_date&  3.9e-2 &-0.06\\
         \hline
    \end{tabular}
    \label{feature_ana}
\end{table}

\begin{figure}[!ht]
    \centering
    \includegraphics[width=\linewidth]{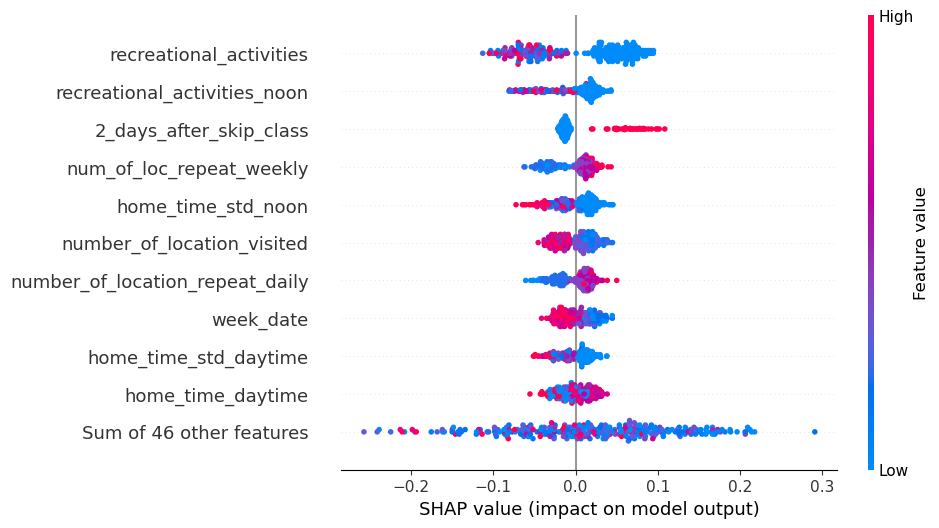}
    \caption{The SHAP beeswarm summary visualization for the stressed classification outcome in the RF model under the AF scenario.}
    \label{fig:XAI}
\end{figure}

Table \ref{feature_ana} demonstrates the substantial predictive power of the proposed features, quantified by significant F-test p-values and their associated correlation coefficients (r-values). Notably, features such as travel time and recreational activity duration, which were extracted via DBSCAN and previously identified as pertinent in Section \ref{rw}, exhibit a robust relationship with stress levels. Our analysis indicates that working time is positively associated with enhanced mental well-being, likely because it provides opportunities for social interaction and a respite from academic demands. Similarly, participation in recreational activities demonstrates a clear beneficial effect on student mental health. Conversely, increased travel time correlates with heightened stress rather than improved well-being. We hypothesize that chronic commuting may induce fatigue \cite{morris2015we}, thereby negatively impacting the volunteers' mental health. 

\begin{figure}[!ht]
    \centering
    \includegraphics[width=0.8\linewidth]{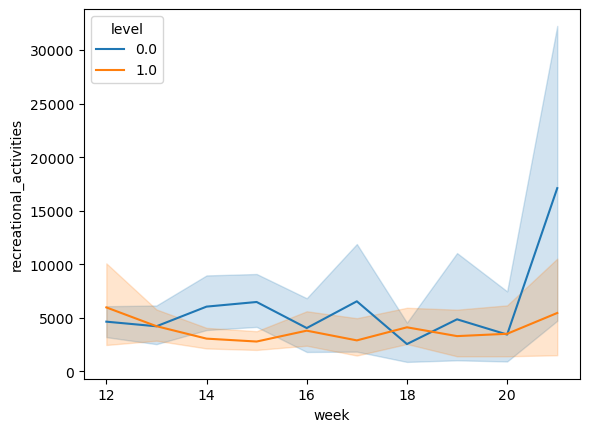}
    \caption{Line plot comparing the recreational activity time (in seconds) of stressed and non-stressed students by week, utilizing PA features. Stressed students are represented by level 1, while non-stressed students are level 0.}
    \label{fig:ra}
\end{figure}

Furthermore, the 'deadline' feature reveals a direct and escalating impact on perceived stress as assignment due dates approach. This observation carries valuable pedagogical implications, suggesting that educators could strategically manage and space out deadlines to mitigate student anxiety and encourage more consistent engagement with academic tasks.

To validate feature importance and enhance model interpretability, we employed Explainable AI (XAI) techniques. As illustrated in Figure \ref{fig:XAI}, the XAI analysis largely corroborates our statistical findings, with six of the top ten features identified by the model aligning with the most statistically significant features from the F-test. The SHAP summary plot effectively visualizes the discriminative power of these variables, showing how fluctuations in feature values correlate with different stress states. For instance, recreational activity duration exhibits a clear divergence between high and low values (represented by red and blue points in Figure \ref{fig:XAI}), strongly influencing the stress classification. This underscores the efficacy of these behavioral markers for privacy-aware stress assessment.

Interestingly, while "time spent at home" did not rank among the top ten features in the F-test, it emerged as highly influential in the XAI analysis. A closer examination reveals that a high standard deviation in home time, indicative of frequent transitions between the residence and other locations, is strongly associated with elevated stress. This high variability likely reflects an erratic pattern of out-of-home activities, which demands additional time and energy, ultimately contributing to fatigue and the accumulation of negative affect. Such findings suggest that irregular residential patterns serve as a crucial complementary marker in modeling stress.
\begin{figure}[!ht]
    \centering
    \includegraphics[width=0.8\linewidth]{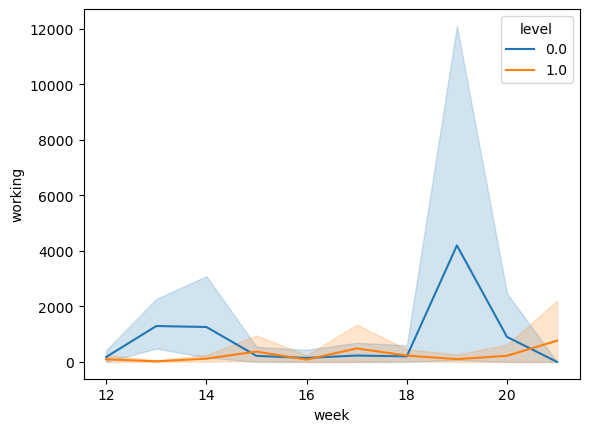}
    \caption{Line plot comparing the workplace time (in seconds) of stressed and non-stressed students by week, utilizing PA features. Stressed students are represented by level 1, while non-stressed students are level 0.}
    \label{fig:w}
\end{figure}

% \subsection{Explainable Capability of PA model}
% As the PA model utilizes a location data stream for human-friendly features extraction, we further conduct a qualitative experiment on the PA feature and its explainable alignment to the current mental health theory.

To further interpret the Privacy-Aware (PA) features, we conducted a temporal analysis aligning behavioral patterns with established mental health literature. As shown in Figure \ref{fig:ra}, non-stressed students consistently spend more time on recreational activities compared to their stressed peers. Specifically, non-stressed individuals average approximately two hours of recreation per outing, roughly 50\% more than those experiencing stress. Furthermore, these students exhibit peaks in recreational time leading up to examination weeks (weeks 18 and 22). This behavioral pattern supports the XAI findings (Figure \ref{fig:XAI}) and aligns with existing literature on the protective effects of leisure against academic pressure \cite{zhang2017academic,chen2025city}. It suggests that allocating time for out-of-home recreation serves as a proactive coping strategy to alleviate pre-exam anxiety and maintain overall mental well-being.

Moreover, Figure \ref{fig:w} highlights a distinct divergence in workplace attendance. Stressed students were rarely observed at a workplace, whereas their non-stressed counterparts consistently spent approximately one hour there during a typical study week. This finding is consistent with Dundes and Marx's observations regarding student employment and well-being \cite{dundes2006balancing}. We hypothesize that the workplace functions as more than just a professional environment; it likely serves as a secondary social hub where students can cultivate meaningful connections, diversifying their daily routines and buffering against academic isolation.

\section{Large-scale Privacy Ablation}\label{ablation_study}
\subsection{Dataset}
\begin{figure}
    \centering
    \includegraphics[width=1\linewidth]{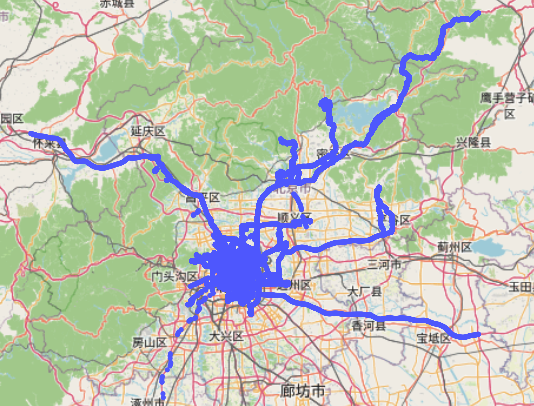}
    \caption{A map visualization of GPS trajectories for 10 random participants in the Beijing GeoLife dataset.}
    \label{fig:Beijing_map}
\end{figure}

To further validate our method's privacy robustness, we performed an additional re-identification attack experiment on the GeoLife dataset \cite{zheng2009mining,zheng2008understanding,zheng2010geolife}. This dataset covers the GPS trajectory data of 182 users between April 2007 and August 2012, spanning more than 1.2 million kilometers and 50,000 hours of data (see Figure \ref{fig:Beijing_map}). In addition, another key strength of the GeoLife dataset lies in its fine-grained GPS data collection, sampled at 1-5 second intervals or every 10 meters. This high resolution allows us to more rigorously evaluate the robustness of our method.

While the dataset has a global reach, the majority of the data was collected in Beijing, China. Our proposed framework is inherently designed for localized, geographically bounded deployments, such as a university campus where a student cohort spends the majority of its time, utilizing a lightweight, self-hosted OSM engine. To accurately simulate this dense spatial environment and accommodate the storage constraints of a localized reverse-geocoding server, we restricted our ablation study to the Beijing subset. Specifically, we filtered the dataset based on the following constraints:

\begin{enumerate}
\item The trajectory data must be located within Beijing.
\item Each user must have more than 18 hours of active GPS data per day.
\item The selected users must meet the above constraints for a duration of more than 5 days.
\end{enumerate}

Ultimately, we identified 83 participants whose data met these criteria for the experiment and .
\subsection{Methodology}

As the GeoLife dataset consists exclusively of GPS trajectories and lacks the multi-modal academic metadata required for the All-Features (AF) set, this ablation study strictly evaluates the purely GPS-driven features configurations.

Therefore, we implemented the raw GPS feature extraction method and the location encoding strategy described in Subsection \ref{LEM}. To evaluate the robustness of various location representations, specifically raw types versus semantic categorizations, we utilized the following raw GPS feature (see Subsection \ref{LEM}) and three configurations for location abstraction, namely:

\begin{itemize}
\item Raw Location Type: This configuration serves as a high-fidelity benchmark, simulating a realistic re-identification attack in a real-world setting where an adversary has access to precise coordinate data.
\item 7-Category Location Grouping: This represents our Privacy-Aware (PA) setting, utilizing broad semantic generalization to mitigate re-identification risks.
\item 12-Category Location Grouping: This configuration of PA quantifies how reduced abstraction affects user privacy, paving the way for new sensitive data transformation methods that promote minimized processing in public datasets. In this work, the 12 categories include: Home, Education, Workplace, Essential Shops (Necessities), Leisure Retail (Shopping), Food \& Drink, Recreation \& Sport, Culture \& Tourism, Infrastructure \& Transit, Roadways (Movement), Health \& Public Services, Miscellaneous \& Utilities.
\end{itemize}

\subsection{Ablation Result on Re-identification Attack}
\begin{table}[!ht]

    \caption{The re-identification attacks accuracy (top 1 accuracy (balanced accuracy)) in 4 scenarios in the GeoLife dataset (in percent). }
    \begin{tabular}{ccccc}
    \hline
        & \textbf{7-Category Location} & \textbf{12-Category Location}& \textbf{Raw Location Type} & \textbf{Raw GPS}\\
        \hline
        \textbf{Rich Knowledge} & 46.97 (24.68)  & 51.74 (31.56) & 99.27 (96.32)&63.30 (51.72)\\
       \textbf{Moderate Knowledge}  & 45.19 (26.08)  & 49.52 (29.00) & 99.41 (96.65)&60.53 (43.74)\\
       \textbf{Limited Knowledge}  &35.69 (14.87)  & 40.14 (17.53) & 96.06 (81.92)&51.24 (32.37)\\
       
    \hline
    \end{tabular}
    
    \label{tab:reid2}
\end{table}
Table \ref{tab:reid2} presents the re-identification accuracy metrics for the GeoLife dataset across four distinct data representation scenarios. The experimental results demonstrate that using Raw Location Types provides an adversary with near-perfect re-identification capabilities, achieving accuracy rates exceeding $99\%$ in both Rich and Moderate knowledge scenarios. Interestingly, in the Raw Location Type configuration, the Moderate Knowledge attack marginally outperformed the Rich Knowledge attack; however, both represent a critical privacy failure.

The findings strongly validate that Privacy-Aware (PA) location transformations, specifically 7-category and 12-category groupings, significantly enhance privacy when processing GPS data. More notably, this experiment highlights a viable path for secure data sharing via pre-defined location grouping. While an attacker can transform raw GPS coordinates into raw location types to achieve nearly $100\%$ re-identification success, the proposed semantic pre-grouping method manages to reduce this top 1 success rate by approximately two to three times and balance accuracy by up to 5 times. This effectively force-multiplies the "anonymity set" and increases the computational burden on the adversary.

\begin{figure}
    \centering
    \includegraphics[width=1\linewidth]{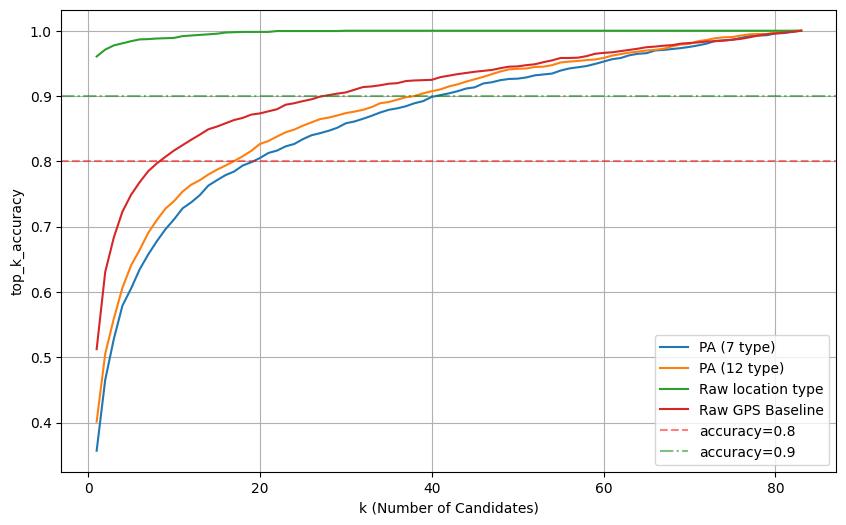}
    \caption{The top-k-accuracy of the re-identification attack in the \textit{Limited Knowledge} scenario for GeoLife dataset}
    \label{fig:k_acc_geolife}
\end{figure}

Furthermore, the Top-$k$ accuracy results for the re-identification attack highlight the robust privacy-preserving characteristics of our proposed system. As illustrated in Figure \ref{fig:k_acc_geolife}, the Privacy-Aware (PA) configurations, including both the 7-category and 12-category models, consistently demonstrate superior privacy guarantees compared to both the Raw GPS and Raw Location Type baselines.

Specifically, an adversary utilizing the PA method requires a candidate set of $k \approx 18$ to achieve an 80\% re-identification accuracy threshold. 
To reach a 90\% accuracy level, the required set expands to $k \approx 40$, which represents a significant increase in the anonymity set (approximately 10 additional candidates) compared to the Raw GPS method. These results further validate the effectiveness of our semantic generalization approach in creating a privacy-aware, high-utility computing framework that fundamentally complicates re-identification attempts.

% \section{Formative User Study}
% \subsection{Survey Collection Procedure}
% To evaluate user perceptions of privacy and explainability, we conducted a formative online survey (N=19). The participants were recruited via convenience sampling, primarily targeting university students to align with the demographic profile of the StudentLife benchmark dataset used in our computational evaluation. Firstly, all participants were informed of the study's academic purpose and provided explicit consent. Moreover, in strict adherence to the privacy-preserving ethos of our proposed framework, the survey was completely anonymous by design. No demographic metadata (e.g., age, gender, or institutional affiliation) was collected. This intentional data minimization strategy was employed to eliminate social desirability bias and encourage highly candid feedback regarding sensitive location-tracking preferences. 
% \subsection{How Human Perceived Privacy and PA method}

% \subsection{How Explainability was perceived}

\section{Discussion}
Contrasting to prior work on grouping the location into the top-k-visit location, a key contribution of this research is the introduction of a novel GPS encoding and LLM-assisted human-understandable feature extraction pipeline. This approach significantly improves upon existing methods, outperforming the GPS-only state-of-the-art \cite{harit2024monitoring} by 6\% in F1 score and accuracy.  Critically, our study also proposes a pathway for leveraging location characteristics to advance stress recognition research in a privacy-conscious manner. By focusing on inherent location-based features, this work paves the way for future studies that prioritize user privacy and the generalizability aspect in mental well-being detection.

\subsection{Position Abstraction and Privacy Protection}
In the realm of ubiquitous sensing, privacy remains an enduring challenge. Our study re-validates the re-identification risks inherent in raw GPS and POI (Point of Interest) scenarios within the StudentLife and GeoLife dataset (see Section \ref{mainresult} and Section \ref{ablation_study}), serving as a cautionary note against the release or use of raw POI/location name data in current research practices \cite{busso2025diversityone,morshed2019prediction,zhang2026narrativesense}.

As an alternative pathway, this study proposes a method that utilizes a self-hosted reverse-geocoding system. By doing so, we prevent sensitive location data from being transmitted to third-party APIs. Instead, only generalized location types are sent to the Large Language Model (LLM), mapping specific functions to pre-defined categories. This approach has proven effective in mitigating re-identification attacks. Furthermore, this method can be scaled into a data-sharing protocol where information is pre-extracted into standardized groups, allowing data providers to share customized subsets based on specific research inquiries. We expect this framework to foster secure data sharing and customization, minimizing privacy risks while maintaining the overall utility of the dataset.

\subsection{Scalability and Maintenance of the Local Infrastructure}\label{scalability}
In this study, we proposed a self-hosted reverse-geocoding server. While our evaluation datasets primarily span specific regions (such as Dartmouth/Hanover or Beijing), this localized geographic focus aligns perfectly with the inherent mobility patterns of our target demographic: university students. Students predominantly spend their time within a bounded geographic ecosystem, circulating between campus facilities, local residences, and nearby recreational or commercial areas.

This regional concentration strongly justifies the implementation of a localized, self-hosted location engine. Moreover, our framework can be seamlessly deployed on standard, institution-owned servers without the need for specialized, high-capacity hardware. This architectural choice makes the system highly scalable for practical, cohort-level deployments. Universities or mental health researchers can adopt this framework to monitor student well-being across an entire campus population without incurring the massive computational costs of global spatial databases, all while maintaining strict data sovereignty by bypassing third-party APIs entirely.

Furthermore, the regional nature of student mobility significantly streamlines the maintenance of the LLM-bootstrapped semantic map. The variety of location types within a typical university town is naturally constrained. Therefore, managing the static semantic dictionary requires minimal overhead. When a student visits an unforeseen location, the system can resolve the edge case via a dynamic LLM call with minimal latency, as the volume of these edge cases is kept low by the cohort's routine behaviors. Institutions can easily sustain this system longitudinally by periodically updating their local dictionaries, ensuring robust, privacy-preserving stress recognition that scales effortlessly to the size of their student body.

\subsection{Socio-Technical Implications and Risks}
Rising academic and professional pressures in the modern world are leading to an increase in mental health challenges, reflecting a worldwide issue where stress negatively impacts both individual well-being and overall productivity. Our work can be applied to address this by allowing users to identify early indicators of mental distress with privacy aware approach. This would empower students to understand more about themselves.

For students experiencing stress, our application features can provide deeper insights into the specific causes of their difficulties. A student may, for example, be struggling to balance work, studies, and their personal life. By leveraging the invaluable transparency machine learning algorithm provides, the student can better understand which factors might be contributing to the issue.

By contrast to this potential, there is a significant risk in the application. A university administration, for instance, could misapply stress-level data in a punitive manner, such as flagging students as "at-risk." This could trigger a cycle of mandatory, and potentially unwelcome, meetings that could paradoxically increase the very stress the system was designed to detect.

% \subsection{Implications for the Ubiquitous Computing Community}
% Beyond these social implications, the PA (Privacy aware) feature extraction pipeline can be integrated into the location data transformation workflow prior to deployment. This method converts raw spatial data into semantic-level locations, enabling future research to more precisely analyze the intersection of daily routines and mental health. This integration is a critical advancement, as it directly addresses the utility-explainability-privacy trilemma, a challenge frequently overlooked in current literature. By navigating this trilemma, the pipeline bridges the gap between interview-based research and the successful deployment of passive mental health assessment applications in real-world environments.

\subsection{Limitations and Future Work}
\subsubsection{Dataset and Deployment Limitations}

The selection of the StudentLife dataset \cite{student_life} represents a deliberate methodological decision to address the Privacy-Utility-Explainability trilemma. To rigorously quantify the effectiveness of our Privacy-Aware (PA) framework, it is imperative to begin with un-obfuscated, raw GPS trajectories---a data component that has become increasingly scarce in modern open-source repositories due to heightened privacy regulations.

While we acknowledge that the intervening decade has introduced shifts in student lifestyle patterns and technology integration, StudentLife remains the canonical benchmark for mobile sensing. It provides the unique, high-fidelity ground truth necessary to demonstrate how raw geographic identifiers can be successfully destroyed without sacrificing behavioral utility. Consequently, while specific behavioral correlations may be temporally bound to the 2014 academic context, our findings regarding the robustness of the privacy-utility trade-off remain methodologically sound and broadly applicable to the next generation of GPS transformations.
% Due to the scarcity of open-source datasets containing raw GPS trajectories required for our baseline comparison, the StudentLife dataset \cite{student_life}, which was collected in 2014, was chosen as a benchmark dataset. Although this dataset is a foundational benchmark for mobile sensing research, the intervening decade has seen some shifts in student behavior, technology integration, and lifestyle patterns. Consequently, while our findings on the privacy-utility trade-off are methodologically robust, the specific behavioral explanations, such as the correlations between stress and time spent at work or recreation, may be temporally bound and not fully representative of current student experiences.

Furthermore, the 20-minute sampling interval in the StudentLife dataset results in relatively coarse location data. Future research should transition to finer intervals, such as one or five minutes, to gain a more granular understanding of human activity and behavioral patterns.
Perhaps more critically, the correlations found between workplace presence and mental health in this study are limited to a "short-duration" group (those working less than two hours per day). In a real-world academic setting, students often work up to 4-6 hours per day, a level of intensity that frequently leads to stress or burnout.

Additionally, since this study utilizes the StudentLife and GeoLife datasets, it adheres to a simulation-based research; no physical system was deployed in a real-world environment. Future research can bridge this gap by implementing this approach within a live data collection system. This would allow for a deeper understanding of practical deployment constraints---such as power consumption and memory limitations---that may influence the scalability and adoption of this method.

\subsubsection{Location Ambiguity and User Context}
Another limitation of our approach lies in the classification of location types. Given the inherently ambiguous function of many locations, our single-label classification poses a potential challenge. For example, a single coffee shop could be a venue for recreation, a remote workplace, or simply a place for dining. This multifunctional characteristic means our system may not capture the user's true context. 

% While this work provides a complete algorithmic framework for privacy-aware stress recognition from data-at-rest, we have also identified the critical path to solving end-to-end privacy.
Thus, future work should aim for a more precise grouping of location functions by aligning both the location type and temporal patterns. A more in-depth analysis could, for instance, use the time of day or day of the week to infer whether a visit to a location is for work-related activities or leisure. 
% More importantly, future studies should also explore deeper in on-device LLM, such as Llama or Qwen, to further enhance users' privacy, location encoding accuracy, and edge-case handling.

\subsubsection{Privacy Scope and Transition Vulnerabilities}
 The privacy safeguards in this research currently rely on the pre-extraction of total time spent in specific location types. While effective for this study, this approach overlooks the temporal sequences inherent in real-world human mobility. Future work should expand upon this pre-extraction method to analyze how location transitions and timing specifically impact the system's privacy profile and mental health recognition.

\section{Conclusion}

In this paper, we present an approach to enhance privacy and compare the performance of stress recognition between the proposed and the traditional method. Our methodology incorporates a GPS encoding strategy, which transforms GPS coordinates into more private location features, thereby ensuring individual privacy. Additionally, we leverage the capabilities of Gemini 2.5 Flash, a Large Language Model (LLM), to categorize these location features into meaningful categories of home, school, recreational activity, shopping, working, travel, and others, which are crucial for assessing students' daily stress status.

Our experimental results demonstrate the effectiveness of our approach. In StudentLife dataset, achieving a 64\% F1 score and 67\% accuracy for the PA setting for the stress level prediction task while decrease top 1 re-identification accuracy by 5 to 20\% compared to AF scenario. In addition, the LOSO scenarios demonstrate the robust generalizability of this approach. On entirely unseen subjects, our Privacy-Aware (PA) model achieved an F1-score of 70\%, a performance statistically indistinguishable from the substantially less private AF model (71\% F1).

In GeoLife experiment, our method also illustrates strong re-identification mitigation from 2 to 3 times and by 5 times balance accuracy re-identification attacks. This not only underscores the potential of GPS data in stress recognition but also highlights the importance of the comprehensive utilization of limited data sources and users' privacy.

Most notably, this study introduces a GPS transformation framework designed for sharing behaviorally meaningful datasets. This 'PA framework' provides researchers with an alternative methodology for data sharing and feature exploration through the use of location data.
% To summary, this research contributes a significant advancement in the application of machine learning techniques for stress recognition, offering a practical and privacy-sensitive approach. The implications of our findings are particularly relevant in educational settings, where understanding and mitigating student stress is crucial.

\begin{acks}
We acknowledge Ho Chi Minh City University of Technology (HCMUT), VNU-HCM, for supporting this study. 
% We also express our gratitude to Christina Garcia, Hong Tri Nguyen, Hoang Nhat Khang Vo, Anh Vy Ngo, and Khuong Cong Duy Nguyen for their valuable comments. Furthermore, we thank the anonymous reviewers for their insightful comments which improved the comprehensiveness of this work. In addition, we are deeply grateful to the anonymous volunteers who participated in our survey. Finally, the Google Gemini LLM was utilized for grammatical refinement and as a feature extractor within this research.
\end{acks}

\bibliographystyle{ACM-Reference-Format}
\bibliography{refs}

\newpage
\appendix
\section{Stress Recognition Ablation studies}
\subsection{Ablation Study Experiments}

To isolate the predictive power of different feature groups, we designed an ablation study. In addition to the full PA (Privacy-Aware) and AF (All-Features) models, we evaluated three configurations using reduced feature sets:

\begin{itemize}
    \item Location-Function Only (LF): A model trained using only the GPS-derived location-function features (e.g., time at home, time at school).
    \item Academic-Only (AO): A model trained using only the academic-based features (e.g., deadlines, class attendance).
    \item LF + AO (LFAO): A combined model using both the Location-Function and Academic-Only features. This configuration effectively represents the PA model without its time-based features.
\end{itemize}

\begin{table}[!ht]
    
        \caption{The extended table of feature type and quantities in each evaluation setting.}
    \begin{tabular}{ccccccccc}
    \hline
     \textbf{Data stream}& \multicolumn{2} {c} {\textbf{Feature type}}  &\textbf{Quantities}  &\textbf{AF}  & \textbf{PA}& \textbf{LF}& \textbf{AO}& \textbf{LFAO}\\
       \hline
     \multirow{3}{*}{\textbf{GPS}}  & \multirow{2}{*}{\textbf{Location}}  & \textbf{Function-based} & 38&  \checkmark & \checkmark&  \checkmark & & \checkmark\\
 & & \textbf{Address-based} & 3& \checkmark &&  &&\\
&\multicolumn{2} {c} {\textbf{Time-based}}&  2 &\checkmark &\checkmark &  & \\
\hdashline
    \textbf{Class and Deadline}   & \multicolumn{2} {c} {\textbf{Academic-based}}&  11 & \checkmark& &   & \checkmark & \checkmark\\

         \hline
        & \multicolumn{2} {c} {\textbf{Number of Features}}&54&54&40&38&11&49\\
         \hline
    \end{tabular}

    \label{AFPA__}
\end{table}

\subsection{Ablation Study Results}
\subsubsection{Re-identification Attack Accuracy}
\begin{figure}[!ht]
    \centering
    \includegraphics[width=\linewidth]{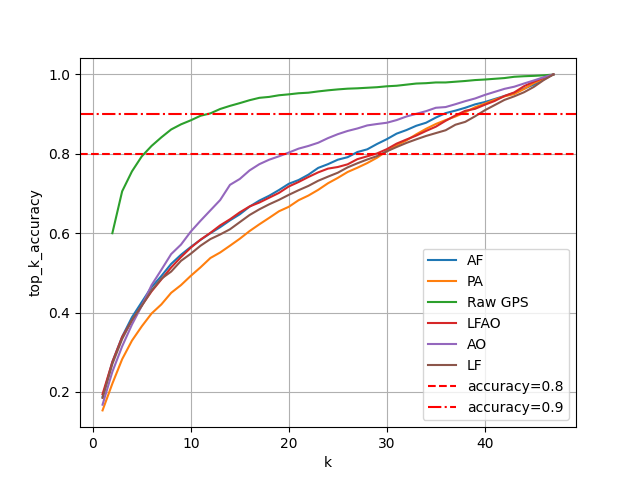}
    \caption{The extended top-k-accuracy of the re-identification attack in the \textit{Limited Knowledge} scenario.}
    \label{ext_PU}
\end{figure}

Figure \ref{ext_PU} illustrates the comprehensive view on re-identification attack on multiple set of features. The Raw GPS baseline is clearly the most vulnerable, achieving 90\% re-identification accuracy with a small candidate pool of only $k \approx 10$. 

Our proposed PA model provides the strongest privacy guarantees, requiring an attacker to amass a significantly larger candidate pool of $k \approx 38$ to reach the same 90\% accuracy. 

The AF model is visibly part of a tight cluster of semantic feature sets, alongside LFAO, LF, and AO. This cluster, while offering a substantial privacy improvement over Raw GPS, is demonstrably more vulnerable to re-identification than the PA model. This, therefore, quantifies the specific privacy risk associated with the high-utility AF model, reinforcing the trade-off that our PA model is designed to solve.

\subsubsection{Stress recognition Accuracy}
The performance of these ablated models was evaluated using the random-splitting method. Table \ref{RS_ablation} compares their accuracy and F1 scores against our primary models and the STGNN GPS baseline.

\begin{table}[!ht]
    \caption{The extended table of classification performance comparison in the random splitting method to the STGNN GPS baseline (in \%). The figure reported was from the best model result from XGB and RF.}
    \centering
    \begin{tabular}{lcccccc}
    \hline
        \textbf{Model} & \textbf{STGNN GPS} & \textbf{PA} & \textbf{AF} & \textbf{LF} & \textbf{AO} & \textbf{LFAO} \\
    \hline
        Acc & 61 & 67 & 68 & 64 & 53 & 64 \\
        F1  & 58 & 64 & 65 & 60 & 52 & 60 \\
    \hline
    \end{tabular}
    \label{RS_ablation}
\end{table}

The results of the ablation study, presented in Table \ref{RS_ablation}, reveal the distinct contributions of each feature group. The AO (Academic-Only) model, relying solely on academic information, achieved the lowest performance (F1 52\%). This indicates that while academic context is a contributing factor, it is insufficient on its own for robust stress recognition. In contrast, the LF (Location-Function) model, which uses only our proposed GPS-derived features, achieved an F1 score of 60\%. This finding is particularly significant, as it demonstrates that our semantic GPS encoding method by itself outperforms the STGNN GPS baseline (F1 58\%), proving its strong standalone predictive power.

The study also highlights a powerful synergistic effect between feature sets. Interestingly, combining location and academic features in the LFAO model (F1 60\%) offered no additional benefit over the LF model alone, suggesting the predictive signals from academic features may be largely captured by the location-function data. The most critical insight, however, is the performance jump from the LFAO model (F1 60\%) to our full PA model (F1 64\%). This 4-point increase is attributed entirely to the inclusion of time-based features. This strongly suggests that temporal context (such as the day of the week) is a crucial component, providing a lens that significantly enhances the model's ability to interpret behavioral and academic patterns for stress recognition.

\section{Location-encoding result}

Table \ref{LLM_Human} presents a comparison between the location types categorized by the Large Language Model (LLM) and those assigned by human labelers. Overall, the LLM classified location types with high accuracy (94/103). A strong correlation was observed between the human labels and the LLM's categories, particularly for the 'Home', 'School', and 'Others' groups. However, the LLM struggled to label recreational places, failing to correctly identify 4 out of 15 locations in this category.

\begin{table*}[!ht]
    
    \caption{The LLM categorizes location type and human labeling. The label in \textbf{bold} is the difference between human and LLM categorization.}
    \begin{tabular}{c|p{0.35\linewidth} p{0.35\linewidth}}
    \hline
    \textbf{Category}&  \textbf{LLM} &  \textbf{Human label}  \\
    \hline
         \textbf{Home}& dormitory, house, apartments, residential & dormitory, house, apartments, residential\\
         \hdashline
         \textbf{School}&university, library, primary, educational\_institution, school, tertiary, college, secondary  &university, library, primary, educational\_institution, school, tertiary, college, secondary  \\
         \hdashline
         \textbf{Travel}& bicycle\_parking, parking, bus\_stop, parking\_space, cycleway, railway\_station, aerodrome, terminal, track, apron  &bicycle\_parking, parking, bus\_stop, parking\_space, cycleway, railway\_station, aerodrome, terminal, track, apron,\textbf{ bridge} \\
         \hdashline
         \textbf{Shop}& retail, florist, supermarket, restaurant, cafe, clothes, bakery, gift, bank, bar, computer, books, jewelry, convenience, electronics, wholesale & retail, florist, supermarket, restaurant, cafe, clothes, bakery, gift, bank, bar, computer, books, jewelry, convenience, electronics, wholesale, \textbf{fast\_food}\\
         \hdashline
        \textbf{Work}&office, estate\_agent, post\_office, hairdresser, interior\_decoration, insurance, telecommunication, chemist, veterinary, beauty, car\_repair, construction, townhall  & office, estate\_agent, post\_office, hairdresser,  interior\_decoration, insurance, telecommunication, chemist, veterinary, beauty, car\_repair,  construction, townhall,\textbf{  industrial, clinic, audiologist} \\
        \hdashline
         \textbf{Recreation}& sports\_centre, bench, artwork, arts\_centre, pitch, museum, outdoor\_seating, grandstand, golf\_course, theatre, pub  & sports\_centre, bench, artwork, arts\_centre, pitch, museum, outdoor\_seating, grandstand, golf\_course, theatre, pub,\textbf{ hotel, picnic\_table,  commercial, courtyard} \\
         \hdashline
         \textbf{Others}& yes, service, place\_of\_worship, courtyard, hotel, post\_box, social\_centre, terrace, vacant, commercial, flagpole, pedestrian, fuel, picnic\_table, unclassified, recycling, pottery, tower, industrial, district, association, social\_facility, fast\_food, footway, phone, church, public\_bookcase, toilets, fire\_alarm\_box, clinic, manhole, bridge, information, shelter, detached, shed, tree, dentist, audiologist  &
         
         yes, service, place\_of\_worship, post\_box, social\_centre, terrace, vacant, flagpole, pedestrian, fuel, unclassified, recycling, pottery, district, association, social\_facility, footway, phone, church, public\_bookcase, toilets, fire\_alarm\_box,  manhole, information, shelter, detached, shed, tree, tower, dentist  \\
    \hline
    \end{tabular}
    
    \label{LLM_Human}
\end{table*}

% \section{Re-identification attacks}
% In ubiquitous computing system, attackers thrives to take advantage on the data is the main concerns for user privacy. This section presents the effort to simulate the attacks and quantify the security of our encoding strategy.
% \subsection{Methodology}
% As the threads model, we designed an attacker who have little, some, and rich knowledge about our users and can sneaked into the database and get the users information.
% In the attack simulation, we compares the encoded dataset (PA and AF) and the raw GPS dataset. In the raw GPS dataset, we extracted five simple statistical features per day (mean, max, min, standard deviation and inter quartile range) of the volunteers latitude and longitude. After extracting AF, PA, and GPS features, we split the data into 3 configuration 80:20, 50:50, and 20:80 train-test to simulate the amount of external dataset that attackers can have. Finally, we trained and validate re-identification XGB model to benchmark our method.

\end{document}